\documentclass[journal,comsoc]{IEEEtran}
\usepackage[T1]{fontenc}
\usepackage{url}
\usepackage{amsthm}
\usepackage{array}
\usepackage{diagbox}
\usepackage{multirow}
\usepackage{verbatim}
\usepackage[usenames,dvipsnames,svgnames,table]{xcolor}
\usepackage[utf8]{inputenc} 
\usepackage{slashbox}

\ifCLASSINFOpdf
  \usepackage[pdftex]{graphicx}

\else

\fi

\usepackage{amsmath}

\usepackage[cmintegrals]{newtxmath}

\usepackage{algorithmic}
\usepackage{algorithm2e}

\ifCLASSOPTIONcompsoc
  \usepackage[caption=false,font=normalsize,labelfont=sf,textfont=sf]{subfig}
\else
  \usepackage[caption=false,font=footnotesize]{subfig}
\fi

\usepackage{fixltx2e}

\definecolor{B}{rgb}{0.2,0.2,0.8}
\definecolor{R}{rgb}{0.8,0.2,0.2}
\definecolor{G}{rgb}{0.0,0.45,0.0}
\definecolor{Blue}{rgb}{0.2,0.2,0.8}
\definecolor{Gray}{rgb}{0.8,0.8,0.8}
\definecolor{Magenta}{rgb}{0.8,0.3,0.4}
\definecolor{Brown}{rgb}{0.6,0.4,0.0}

\begin{document}

\title{ A Reduced Codebook and Re-Interpolation Approach for Enhancing Quality in Chroma Subsampling }
%
%
%

\author{
 Kuo-Liang Chung \\
  Department of Computer Science and Information Engineering\\
  National Taiwan University of Science and Technology\\
  No. 43, Section 4, Keelung Road, Taipei, 10672, Taiwan, R.O.C. \\
  \texttt{klchung01@gmail.com} \\
  $\newline$
  Chen-Wei Kao \\
  Department of Computer Science and Information Engineering\\
  National Taiwan University of Science and Technology\\
  No. 43, Section 4, Keelung Road, Taipei, 10672, Taiwan, R.O.C. \\
}

\maketitle

\begin{abstract}
Prior to encoding RGB full-color images or Bayer color filter array (CFA) images, chroma subsampling is a necessary and crucial step at the server side. In this paper, we first propose a flow diagram approach to analyze the coordinate-inconsistency (CI) problem and  the upsampling process-inconsistency (UPI) problem existing in the traditional and state-of-the-art chroma subsampling methods under the current coding environment. In addition, we explain why the two problems degrade the quality of the reconstructed images.
Next, we propose a reduced codebook and re-interpolation (RCRI) approach to solve the two problems for enhancing the quality of the reconstructed images. Based on the testing RGB full-color images and Bayer CFA images, the comprehensive experimental results demonstrated  at least 1.4 dB and 2.4 dB  quality improvement effects, respectively, of our RCRI approach against the CI and UPI problems for the traditional and state-of-the-art chroma subsampling methods.
\end{abstract}

%
\IEEEpeerreviewmaketitle

\section{Introduction}\label{sec:I}
\begin{figure}[htbp]
  \centering
   \includegraphics[scale=0.38]{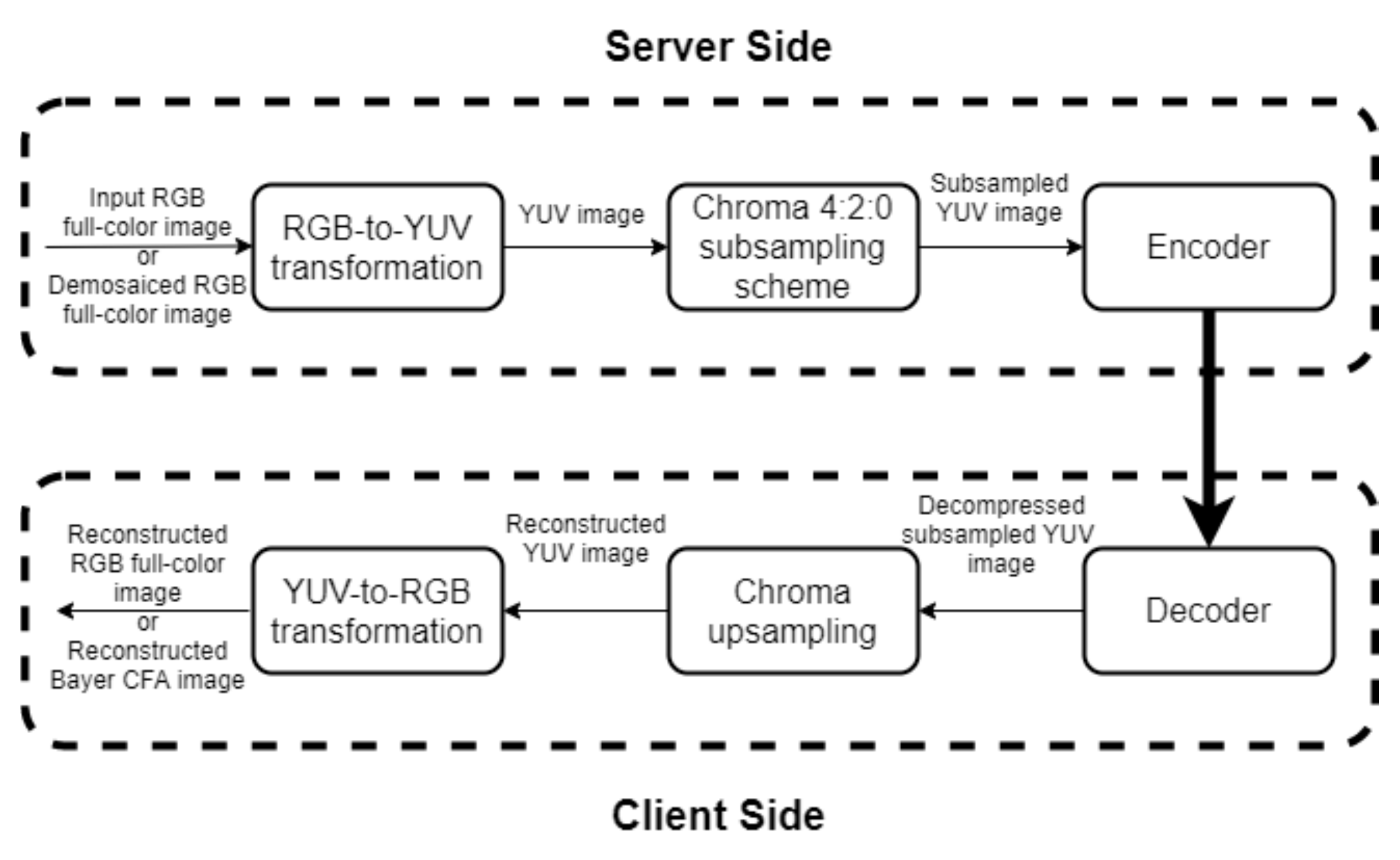}
\caption{The current coding system.}
\label{pic:traditional flow chart}
\end{figure}

%
%
%
%
\IEEEPARstart
{A}{s} shown at the server side of Fig. \ref{pic:traditional flow chart}, in our study, the input image could be a RGB full-color image $I^{RGB}$ or a demosaiced RGB full-color image which has been demosaicked from the input Bayer color filter array (CFA) image $I^{Bayer}$ \cite{B. Bayer}. To demosaick $I^{Bayer}$ to a RGB full-color image, several demosaicking methods \cite{Li-2008},
\cite{Condat-2009}, \cite{Zhang-2011}, \cite{Kiku-2013}, \cite{K. L. Hua}, \cite{W. Ye}, \cite{Y. Monno}, \cite{Z. Ni} can be used; here, the demosaicking method in \cite{Kiku-2013} is used.
For easy exposition, we take the Bayer CFA pattern in Fig. \ref{pic:Bayer pattern}(a) as the representative, but our discussion is also applicable to the other three patterns in Figs. \ref{pic:Bayer pattern}(b)-(d). Prior to compression, according to BT.601-5 \cite{BT.601}, the RGB full-color image is converted to a YCbCr image $I^{YCbCr}$ using the following RGB-to-YCbCr color transformation:
\begin{equation}
  \label{eq:RGB2YCbCr}
  \begin{bmatrix} Y_{i} \\ Cb_{i} \\ Cr_{i} \end{bmatrix} =
  \begin{bmatrix} 0.257 & 0.504  & 0.098 \\
                 -0.148  & -0.291 & 0.439 \\
                  0.439 & -0.368 & -0.071 \end{bmatrix}
  \begin{bmatrix} R_{i} \\ G_{i} \\ B_{i} \end{bmatrix} +
  \begin{bmatrix} 16 \\ 128 \\ 128 \end{bmatrix}
\end{equation}
where for each 2$\times$2 YCbCr block $B^{YCbCr}, (Y_i , Cb_i , Cr_i)$, $1 \leq i \leq 4$, denotes the YCbCr triple-value in zigzag order; $(R_i , G_i , B_i)$ denotes the collocated RGB triple-value of the 2$\times$2 RGB full-color block $B^{RGB}$.

Chroma subsampling has two formats, namely 4:2:0 and 4:2:2. 4:2:0 subsamples the $(Cb, Cr)$-pair for each 2$\times$2 CbCr block $B^{CbCr}$ and 4:2:2 subsamples the $(Cb, Cr)$-pair for each row of $B^{CbCr}$. Throughout this paper, we focus on 4:2:0, although the approach is also is applicable to 4:2:2. 4:2:0 has been used in Bluray discs (BDs) and digital versatile discs (DVDs) for storing movies, sports, TV shows, etc.

After decompressing the encoded subsampled YCbCr image by the decoder, as depicted at the client side of Fig. \ref{pic:traditional flow chart}, a chroma upsampling process is performed on the subsampled CbCr image. Further, the upsampled YCbCr image is converted to a reconstructed RGB full-color image using the following YCbCr-to-RGB color conversion:

\begin{equation}
  \label{eq:YCbCr2RGB}
  \begin{small}
  \begin{bmatrix} R_i \\ G_i \\ B_i \end{bmatrix} =
    \begin{bmatrix} 1.164 & 0 & 1.596 \\ 1.164 & -0.391 & -0.813 \\ 1.164 & 2.018 & 0 \end{bmatrix}
      \begin{bmatrix} Y_i-16 \\ Cb_i-128 \\ Cr_i-128 \end{bmatrix}
  \end{small}
\end{equation}
Suppose the input image is the demosaiced RGB full-color image. By Eq. (\ref{eq:YCbCr2RGB}), the upsampled YCbCr image can be directly converted to the reconstructed Bayer CFA image as the output.

In Subsection I.A, we introduce the related chroma subsampling works for $I^{RGB}$. Then, in Subsection I.B, we introduce the related chroma subsampling works for $I^{Bayer}$.

\subsection{Related Works for $I^{RGB}$}\label{sec:IA}
At the server side, suppose the input image is a RGB full-color image. We first introduce the five traditional chroma subsampling methods, namely 4:2:0(A), 4:2:0(L), 4:2:0(R), 4:2:0(DIRECT), and 4:2:0(MPEG-B) \cite{MPEG-B}. Then, we introduce five state-of-the-art chroma subsampling combinations \cite{Y. Zhang}, \cite{S. Wang}, \cite{K. L. Chung-2020(rgb)}, \cite{Lin-2019}.

\begin{figure}[htbp]
  \centering
    \subfloat[]{
        \begin{minipage}{0.11\textwidth}
          \includegraphics[width=1\textwidth]{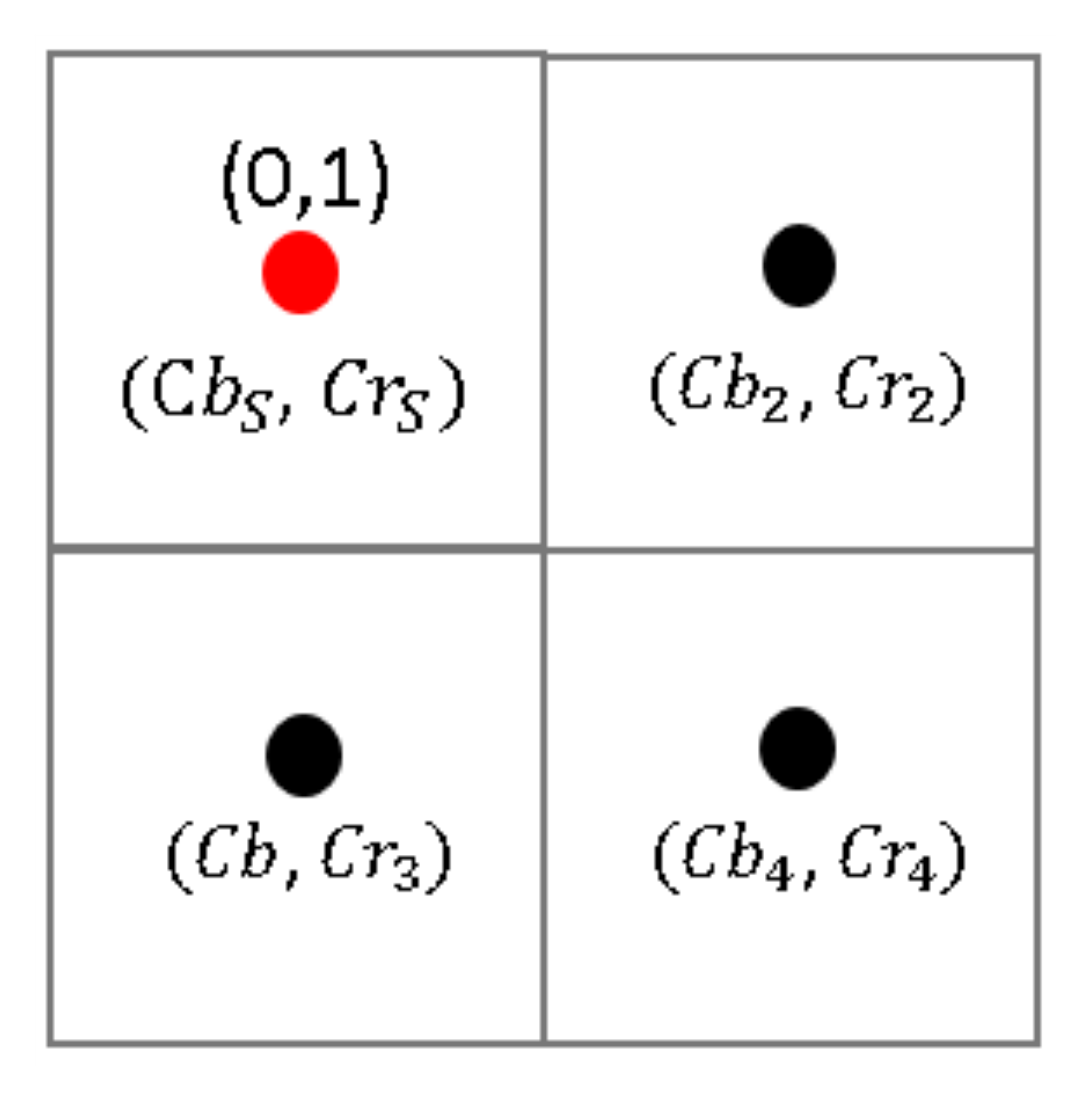}
        \end{minipage}}
    \subfloat[]{
        \begin{minipage}{0.11\textwidth}
          \includegraphics[width=1\textwidth]{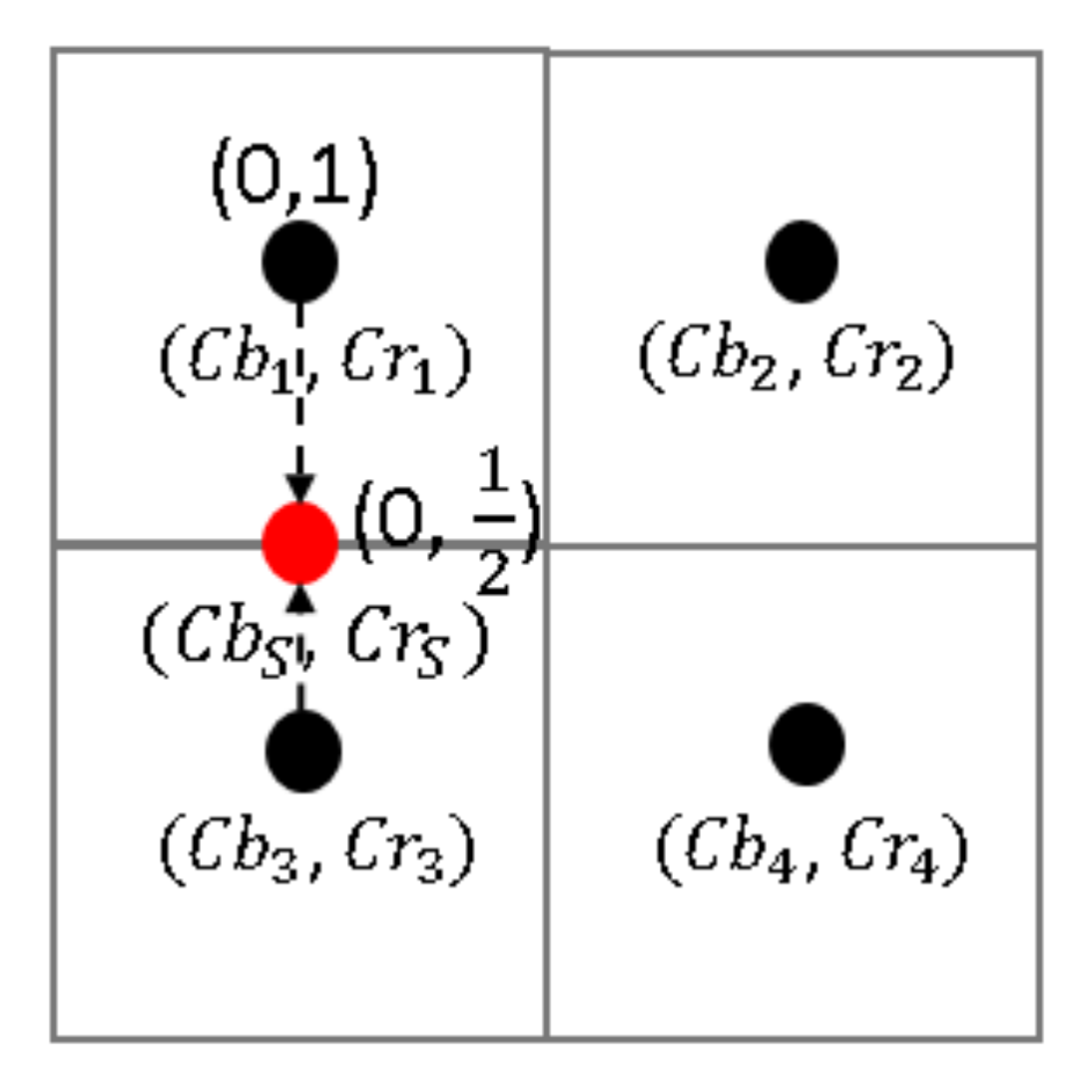}
        \end{minipage}}
    \subfloat[]{
        \begin{minipage}{0.11\textwidth}
          \includegraphics[width=1\textwidth]{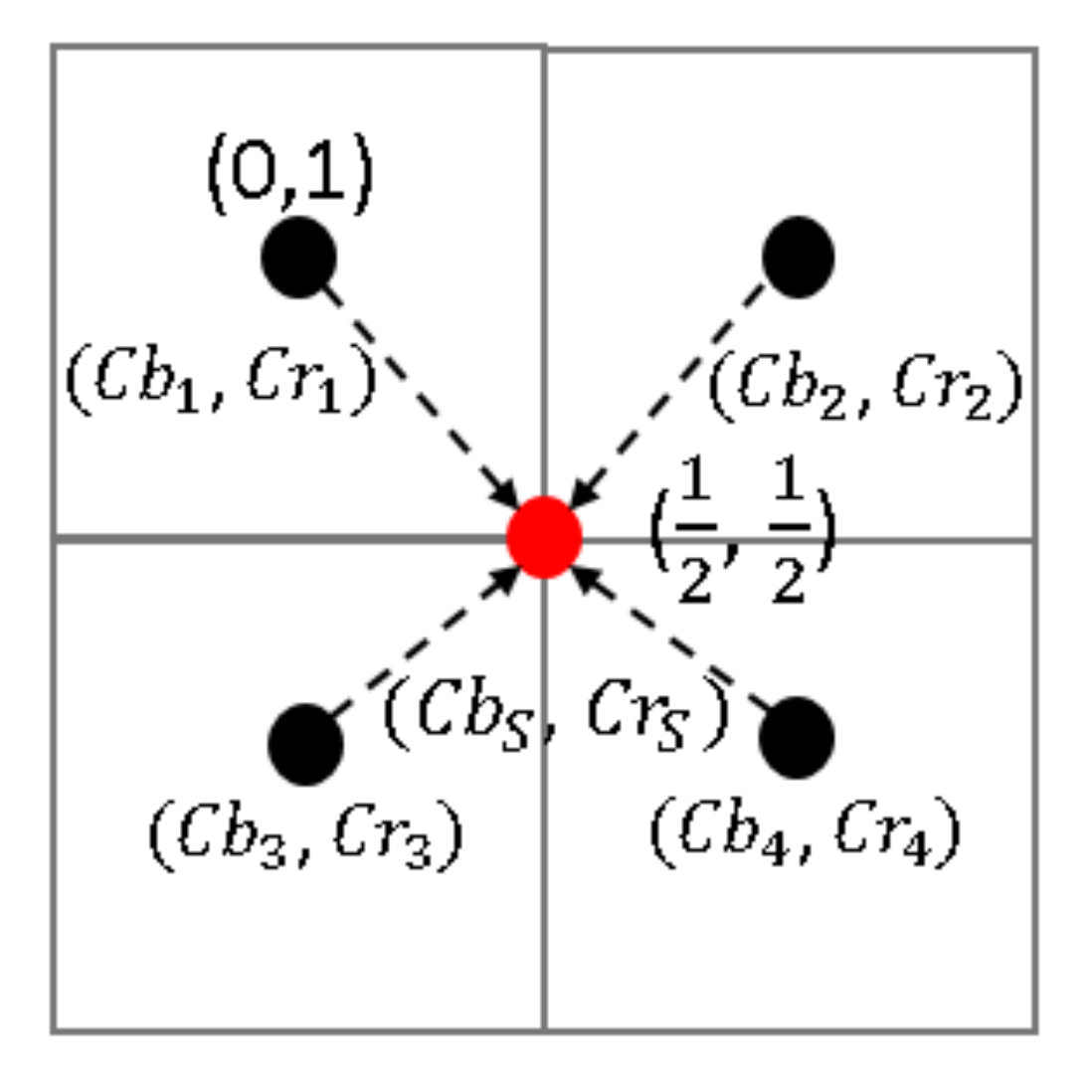}
        \end{minipage}}
    \subfloat[]{
        \begin{minipage}{0.11\textwidth}
          \includegraphics[width=1\textwidth]{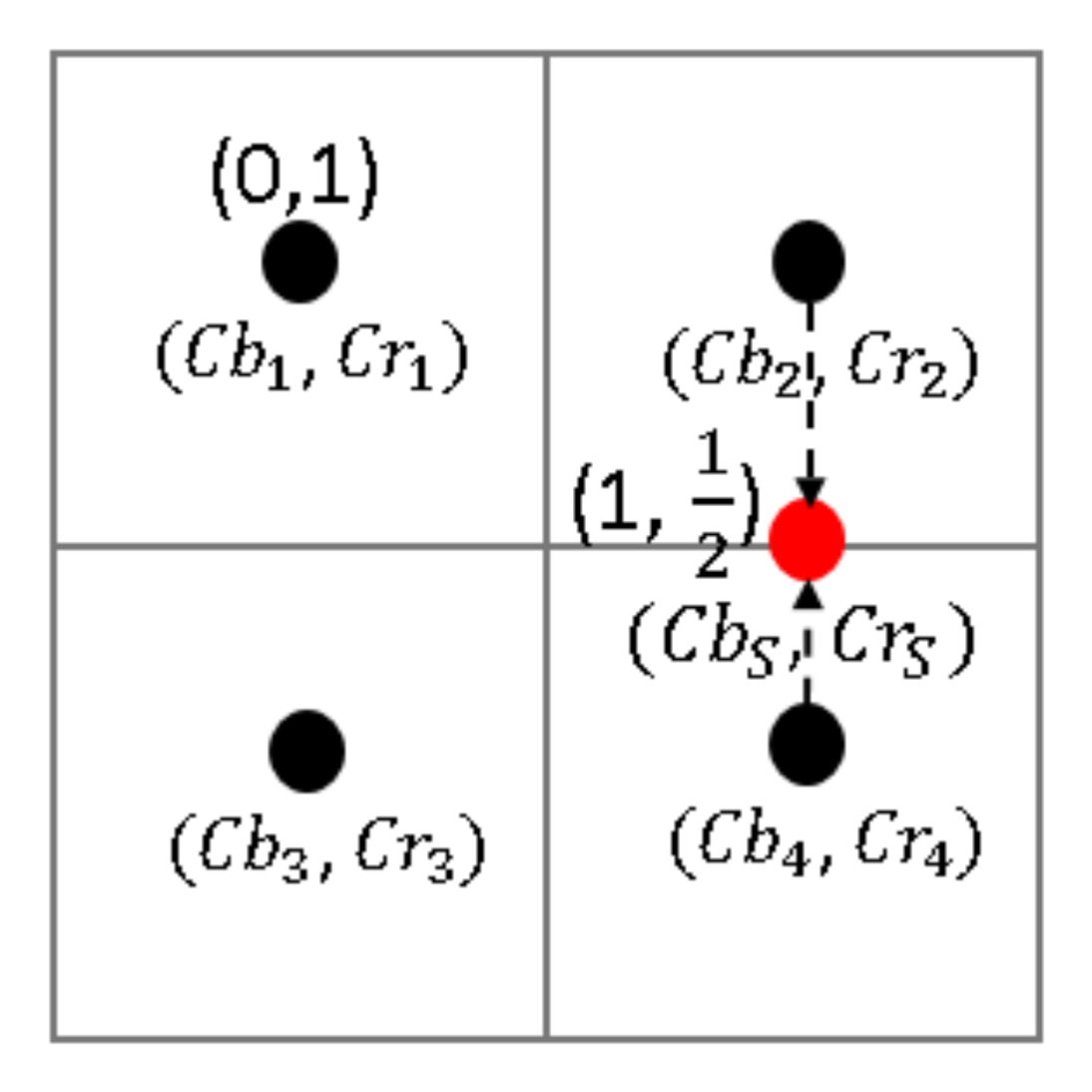}
        \end{minipage}}
\caption{The subsampled chroma positions of the five traditional chroma subsampling methods. (a) For 4:2:0(D) and 4:2:0(MPEG-B). (b) For 4:2:0(L). (c) For 4:2:0(A). (d) For 4:2:0(R).}
\label{pic:subsampling scheme}
\end{figure}

\begin{figure}[htbp]
  \centering
    \subfloat[]{
        \begin{minipage}{0.1\textwidth}
          \includegraphics[width=1\textwidth]{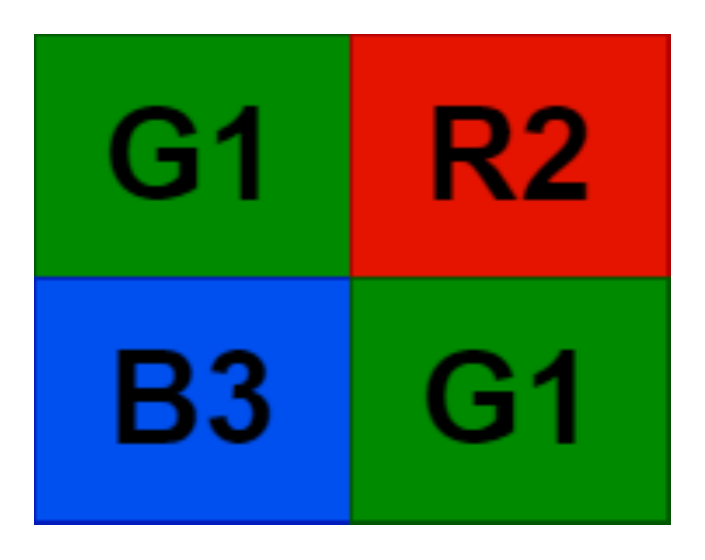}
          \label{GRBG}
        \end{minipage}}
    \subfloat[]{
        \begin{minipage}{0.1\textwidth}
          \includegraphics[width=1\textwidth]{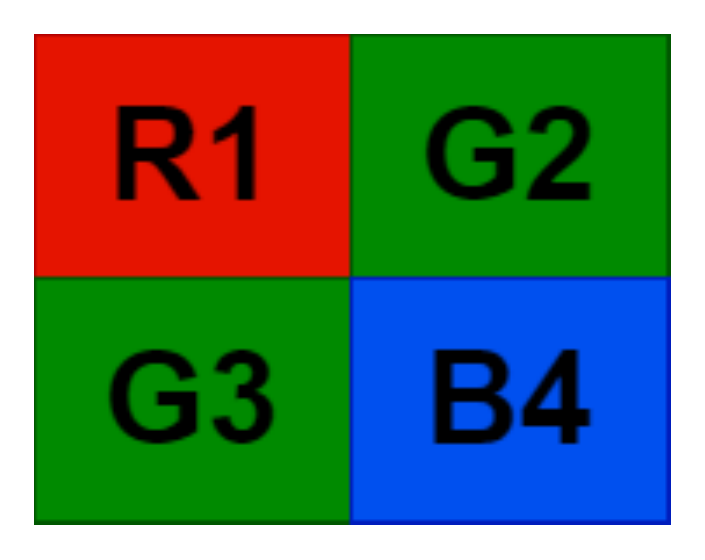}
          \label{RGGB}
        \end{minipage}}
    \subfloat[]{
        \begin{minipage}{0.1\textwidth}
          \includegraphics[width=1\textwidth]{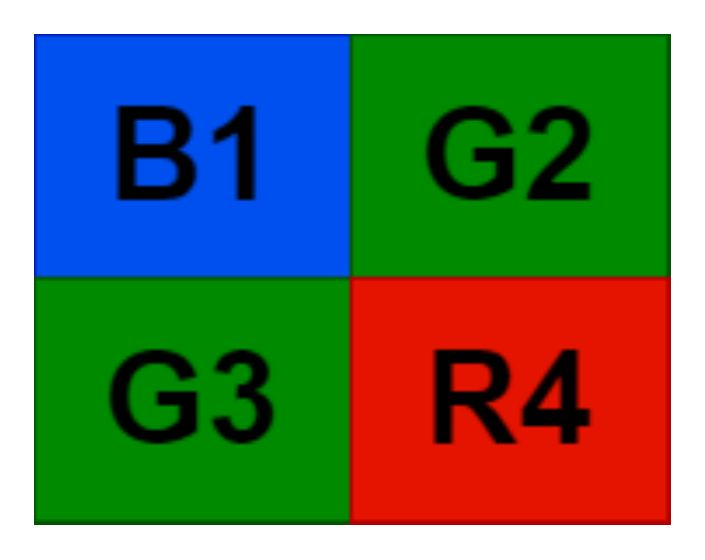}
          \label{BGGR}
        \end{minipage}}
    \subfloat[]{
        \begin{minipage}{0.1\textwidth}
          \includegraphics[width=1\textwidth]{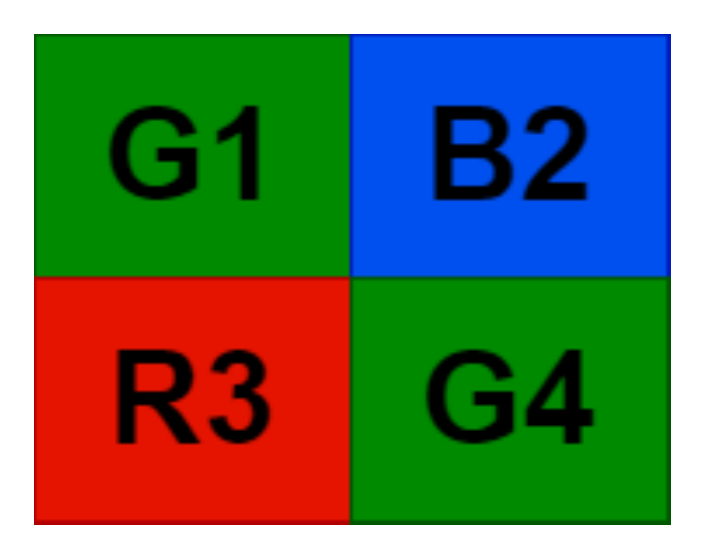}
          \label{GBRG}
        \end{minipage}}
\caption{The four Bayer CFA patterns. (a) $[G_1, R_2, B_3, G_4]$. (b) $[R_1, G_2, G_3, B_4]$. (c) $[B_1, G_2, G_3, R_4]$. (d) $[G_1, B_2, R_3, G_4]$.}
\label{pic:Bayer pattern}
\end{figure}

Among the five traditional chroma subsampling methods, 4:2:0(DIRECT) subsamples the top-left chroma pair of $B^{CbCr}$ as the subsampled $(Cb, Cr)$-pair of $B^{CbCr}$. For simplicity, 4:2:0(DIRECT) is abbreviated as 4:2:0(D). 4:2:0(MPEG-B) determines the subsampled $(Cb, Cr)$-pair by performing the 13-tap filter with mask [2, 0, -4, -3, 5, 19, 26, 19, 5, -3, -4, 0, 2]/64 on the top-left location of $B^{CbCr}$. 4:2:0(L) and 4:2:0(R) subsample the $(Cb, Cr)$-pairs by averaging the chroma components in the left and right columns of $B^{CbCr}$, respectively.
4:2:0(A) subsamples the $(Cb, Cr)$-pair of $B^{CbCr}$ by averaging the four $(Cb, Cr)$-pairs of $B^{CbCr}$.
The subsampled chroma positions of 4:2:0(D) and 4:2:0(MPEG-B) are located at (0, 1); the subsampled chroma positions of 4:2:0(L), 4:2:0(A), and 4:2:0(R)
are located at (0, $\frac{1}{2}$), ($\frac{1}{2}$, $\frac{1}{2}$), and (1, $\frac{1}{2}$), respectively, and the four subsampled chroma positions are marked by
 the four red bullets in Figs. \ref{pic:subsampling scheme}(a)-(d).

According to the new edge-directed interpolation (NEDI) \cite{Li-2001} based chroma upsampling process which improved the previous method \cite{Allebach-1996}, Zhang {\it et al.} \cite{Y. Zhang} proposed an interpolation-dependent image downsampling (IDID) based chroma subsampling method. Their combination is expressed as IDID-NEDI. To improve IDID-NEDI, Wang {\it et al.} \cite{S. Wang} deployed the palette mode \cite{W. Pu} in their JCDU (joint chroma downsampling and upsampling) based chroma subsampling method, and their best combination is expressed as JCDU-BICU, where BICU denotes the bicubic interpolation based chroma upsampling process. The experimental data demonstrated that JCDU-BICU outperforms IDID-NEDI and JCDU-BILI, in which BILI denotes the bilinear interpolation based chroma upsampling process, particularly for screen content images (SCIs) \cite{Y. Lu}.


Following the COPY-based chroma upsampling process and the differentiation technique used in \cite{Lin-2016}, but considering the demosaiced RGB full-color block-distortion as the criterion, Lin {\it et al.} \cite{Lin-2019} proposed a modified 4:2:0(A) chroma subsampling method which selects the best case among the four subsampled $(Cb, Cr)$-pairs of $B^{CbCr}$ by considering the ceiling operation-based 4:2:0(A) and the floor operation-based 4:2:0(A).
Naturally, Lin {\it et al.}'s chroma subsampling method is suitable for the input RGB full-color image. At the client side, they improved the chroma upsampling process \cite{Lin-2019} by considering the distance between each missing chroma value and its three neighboring known (TN) pixels to achieve good quality performance. Their combination is expressed as ``modified 4:2:0(A)-TN''.

Differing from the the chroma subsampling-first luma modification method \cite{K. L. Chung}, in \cite{K. L. Chung-2020(rgb)}, based on the subsampled chroma parameter-pair, at the server side, a BILI-based chroma estimation of $B^{CbCr}$ is deployed in the block-distortion function with two chroma parameters and four luma parameters. Next, using a multiple linear regression technique,
a joint chroma subsampling and luma modification (CSLM) method \cite{K. L. Chung-2020(rgb)} was proposed to determine the subsampled $(Cb, Cr)$-pair of $B^{CbCr}$ and the four modified luma values of $B^{Y}$ simultaneously. Their combination is expressed as CSLM-BILI. Experimental data indicated the quality superiority of CSLM-BILI relative to IDID-NEDI \cite{Y. Zhang}, JCDU-BICU \cite{S. Wang}, and modified 4:2:0(A)-TN \cite{Lin-2019}.


\subsection{Related Works for $I^{Bayer}$}\label{sec:IB}
For $I^{Bayer}$, we mainly introduce the four state-of-the-art chroma subsampling methods \cite{Lin-2016}, \cite{K. L. Chung-2019}, \cite{Lin-2019}, \cite{K. L. Chung-2020(bayer)}.

Chen {\it et al.} \cite{H. Chen} observed that in Eq. (\ref{eq:YCbCr2RGB}), the R value is dominated by the Y and V values, and the B value is dominated by the Y and U values, and then the subsampled $(Cb, Cr)$-pair of $B^{CbCr}$ equals $(U_3, V_2)$ by considering the Bayer CFA pattern. Although their method benefits the R and B components of the reconstructed Bayer CFA image, it does not benefit the G components at all.
To overcome this disadvantage, based on the COPY-based upsampling process to estimate the four chroma-pairs of $B^{CbCr}$ at the server side, Lin {\it et al.} \cite{Lin-2016} proposed a 2$\times$2 Bayer CFA block-distortion function to measure the distortion between the original 2$\times$2 Bayer CFA block and the estimated one. Using a distortion minimization (DM) technique, Lin {\it et al.} derived a closed form to determine the subsampled $(Cb, Cr)$-pair of $B^{CbCr}$. Their combination is expressed as DM-COPY.

In the gradient descent-based (GD) method \cite{K. L. Chung-2019}, the $2\times2$ Bayer CFA block-distortion function used in the DM method \cite{Lin-2016} is proved to be a convex function. Then, according to the shape similarity of the convex function in the real domain to that in the integer domain, an iterative procedure, in which the closed form derived in DM \cite{Lin-2016} was used as the initially subsampled $(Cb, Cr)$-pair of $B^{CbCr}$, was proposed to better improve the DM method. In each iteration, the GD method applied the BILI method to estimate the four $(Cb, Cr)$-pairs of each neighboring chroma block in the eight neighboring chroma blocks, and then the GD method selected the best one with the minimal 2$\times$2 Bayer CFA block-distortion. Their combination is expressed as GD-BILI \cite{K. L. Chung-2019}, and it has better quality than DM-COPY and GD-COPY.

As introduced in the last paragraph of Subsection I.A, Lin {\it et al.}'s modified 4:2:0(A) chroma subsampling method \cite{Lin-2019} only considers the demosaiced RGB full-color block-distortion as the distortion minimization criterion. After performing the modified 4:2:0(A)-TN combination on the demosaiced RGB full-color image, by Eq. (\ref{eq:RGB2YCbCr}), the reconstructed Bayer CFA image can be extracted from the reconstructed RGB full-color image. Therefore, `modified 4:2:0(A)-TN' is also applicable for the input Bayer CFA image.

In \cite{K. L. Chung-2020(bayer)}, based on the BILI-based chroma upsampling process at the server side, combining chroma subsampling, luma modification,  and the Bayer CFA pattern together, a $CSLM^{Bayer}$-BILI combination was proposed.
For each 2$\times$2 YCbCr block $B^{YCbCr}$, $CSLM^{Bayer}$-BILI determined the best solution of the subsampled $(Cb, Cr)$-pair and the modified luma values for $B^{YCbCr}$ simultaneously. In particular, after analyzing all the sixteen (= $2^4$) luma-selection cases, only two luma parameters, namely $Y_1$ and $Y_2$, are modified such that the 2$\times$2 Bayer CFA block-distortion could be minimized.
Experimental data demonstrated that the $CSLM^{Bayer}$-BILI combination \cite{K. L. Chung-2020(bayer)} outperforms DM-COPY \cite{Lin-2016}, GD-BILI \cite{K. L. Chung-2019}, and modified 4:2:0(A)-TN \cite{Lin-2019}.

\subsection{Motivation}\label{sec:IC}
From the introduction of the related chroma subsampling works for $I^{RGB}$ and $I^{Bayer}$, we find that under the current coding system in Fig. \ref{pic:traditional flow chart}, the traditional and state-of-the-art combinations tend to suffer from the coordinate-inconsistency (CI) problem and/or the chroma upsampling process-inconsistency (UPI) problem because at the client side, the decoder is blind to the chroma subsampling process used at the server side and the future chroma upsampling process prefered by the chroma subsampling method. The two problems will be defined in Section III in detail.

The CI and UPI problems lead to the quality degradation of the reconstructed images, which will be explained in Section III. The two problems prompted us to develop a systematic approach to analyze them, and then to propose an effective approach to solve them for enhancing the quality of the reconstructed images.

\subsection{Contributions}\label{sec:ID}
In this paper, we first analyze the subsampled chroma positions of all considered chroma subsampling methods, and then we partition all these chroma subsampling methods into four classes. Based on the four partitioned chroma subsampling classes and the allowable chroma upsampling processes,
we propose a flow diagram approach to analyze the coordinate-inconsistency (CI) problem and the upsampling process-inconsistency (UPI) problem occurring in the traditional and state-of-the-art combinations. We also explain why the two problems lead to the quality degradation of the reconstructed images.

To solve the CI and UPI problems, we propose an effective reduced codebook and re-interpolation (RCRI) approach for enhancing the quality of the reconstructed images. Based on the testing RGB full-color images and Bayer CFA images collected from the IMAX \cite{IMAX} dataset, the Kodak \cite{Kodak} dataset, and the Video \cite{Video} dataset, the thorough experimental results justified the significant quality enhancement effects of our RCRI approach against the CI and UPI problems for the traditional and state-of-the-art chroma subsampling methods. For $I^{RGB}$ and $I^{Bayer}$, the average CPSNR (color peak signal-to-noise ratio) gain and the average PSNR gain of our RCRI approach are at least 1.4 dB and 2.4 dB, respectively.

The rest of this paper is organized as follows. In Section II, based on the subsampled chroma positions, all considered chroma subsampling methods are partitioned into four classes. In Section III, the proposed flow diagram approach is presented to analyze
the CI problem and the UPI problem. In Section IV, the proposed RCRI approach is presented to solve the two problems. In Section V, the thorough experimental results are illustrated to justify the significant quality enhancement using our RCRI approach. In Section VI, some concluding remarks are addressed.

\section{ PARTITIONING ALL CONSIDERED CHROMA SUBSAMPLING METHODS INTO FOUR CLASSES }\label{sec:II}

\begin{figure}[htbp]
    \centering
    \includegraphics[scale=0.55]{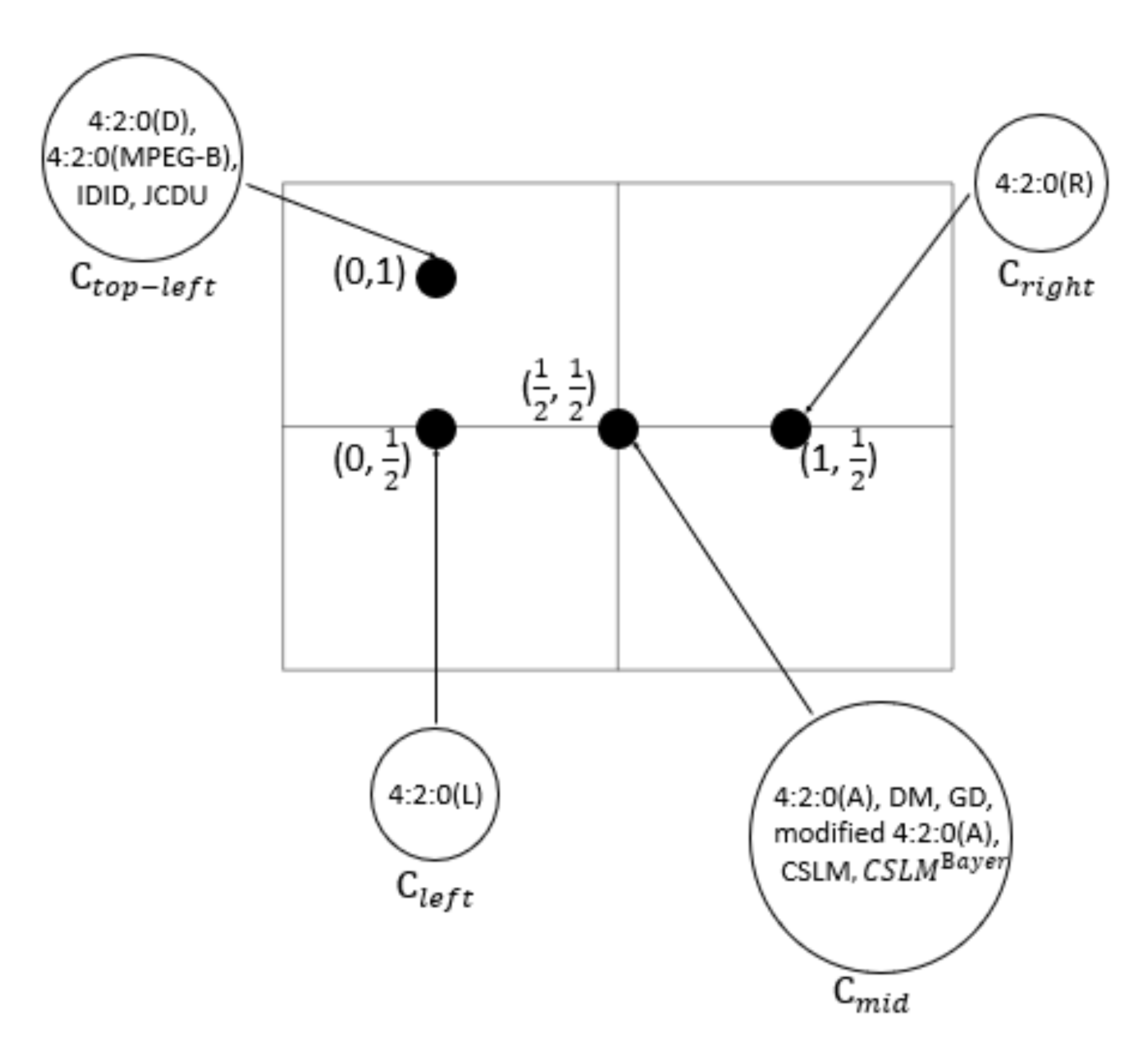}
    \caption{The four subsampled chroma positions for all considered chroma subsampling methods.}
\label{pic:position2}
\end{figure}

In this section, based on the subsampled chroma positions of all considered chroma subsampling methods, we partition these methods into four classes. It has been known that the subsampled chroma positions of the five traditional chroma subsampling methods are depicted by the four red bullets in Figs. \ref{pic:subsampling scheme}(a)-(d) corresponding to the top-left, left, middle, and right black bullets in Fig. 4, respectively. Similar to 4:2:0(D) and 4:2:0(MPEG-B), the subsampled chroma positions of IDID \cite{Y. Zhang} and JCDU \cite{S. Wang} are all located at (0, 1), as depicted by the top-left black bullet of Fig. \ref{pic:position2}.

Based on the subsampled chroma parameter-pair, ($Cb_s$, $Cr_s$), which is computationally located at ($\frac{1}{2}$, $\frac{1}{2}$), the DM \cite{Lin-2016}, GD \cite{K. L. Chung-2019},
and modified 4:2:0(A) \cite{Lin-2019} methods apply the same COPY-based chroma upsampling process to estimate the four chroma pairs of each 2$\times$2 chroma block $B^{CbCr}$ for building up their own block-distortion functions.
Accordingly, the determined values of the subsampled ($Cb_s$, $Cr_s$)-pairs using the above three methods are all located at ($\frac{1}{2}$, $\frac{1}{2}$), as depicted by the middle black bullet in Fig. \ref{pic:position2}. Based on the subsampled chroma parameter-pair located at ($\frac{1}{2}$, $\frac{1}{2}$),
the CSLM \cite{K. L. Chung-2020(rgb)} and $CSLM^{Bayer}$ \cite{K. L. Chung-2020(bayer)} methods apply the BILI-based chroma upsampling process to estimate the four chroma pairs of each 2$\times$2 chroma block $B^{CbCr}$ for building up their own block-distortion functions. Therefore, the subsampled chroma position of the two methods is expressed as ($\frac{1}{2}$, $\frac{1}{2}$).

For all considered chroma subsampling methods, the four subsampled chroma positions (SCPs) are denoted by {\bf SCP} = \{(0, 1), (0, $\frac{1}{2}$), ($\frac{1}{2}$, $\frac{1}{2}$), (1, $\frac{1}{2}$)\}
corresponding to the four partitioned chroma subsampling classes which are expressed as {\bf CS} = \{$C_{top-left}$, $C_{left}$, $C_{mid}$, $C_{right}$\}, as depicted in Fig. \ref{pic:position2}. Accordingly,
in our study, instead of considering the original chroma subsampling method used in one combination, we only consider its chroma subsampling class which the chroma subsampling method used belongs to.

\section{ THE ANALYSIS OF THE UPI AND CI PROBLEMS }\label{sec:III}
We first define the UPI problem and explain why it degrades the quality of the reconstructed color images. Next, we propose a flow diagram approach to analyze the CI problem systematically. According to the coordinate displacement analysis, we explain why the CI problem also degrades the quality of the reconstructed color images.

\subsection{The UPI Problem}\label{sec:ID}
Without the loss of generality, we take the CSLM-BILI combination \cite{K. L. Chung-2020(rgb)} as the example to define the UPI problem.

In CSLM-BILI, as introduced in Subsection I.A, at the server side, a BILI-based chroma estimation of each 2$\times$2 CbCr block $B^{CbCr}$ is deployed in the 2$\times$2 RGB full-color block-distortion function with the parameter-pair, $(Cb, Cr)$.

At the client side, instead of `BILI', if the decoder adopts the other upsampling process, e.g. `COPY', to upsample the received subsampled chroma image, it causes an upsampling process-inconsistency (UPI) problem because the future chroma upsampling process `BILI' preferred by CSLM-BILI is misused as `COPY' by the decoder.

\subsection{The CI Problem}\label{sec:IIIB}
Let the set symbol `{\bf CU}' denote the four chroma upsampling processes, namely COPY, BILI, NEDI, and BICU, used at the client side. Let {\bf CS}x{\bf CU}, where the symbol `x' indicates the cross product operator, denote all combinations over the product of {\bf CS} and {\bf CU}. For each combination in {\bf CS}x{\bf CU}, we propose a flow diagram approach to analyze whether a true CI problem occurs in that combination and to report the coordinate displacement of the true CI problem. Furthermore, we explain why the true CI problem leads to the quality degradation of the reconstructed images.

\begin{figure*}
  \centering
  \includegraphics[scale=0.39]{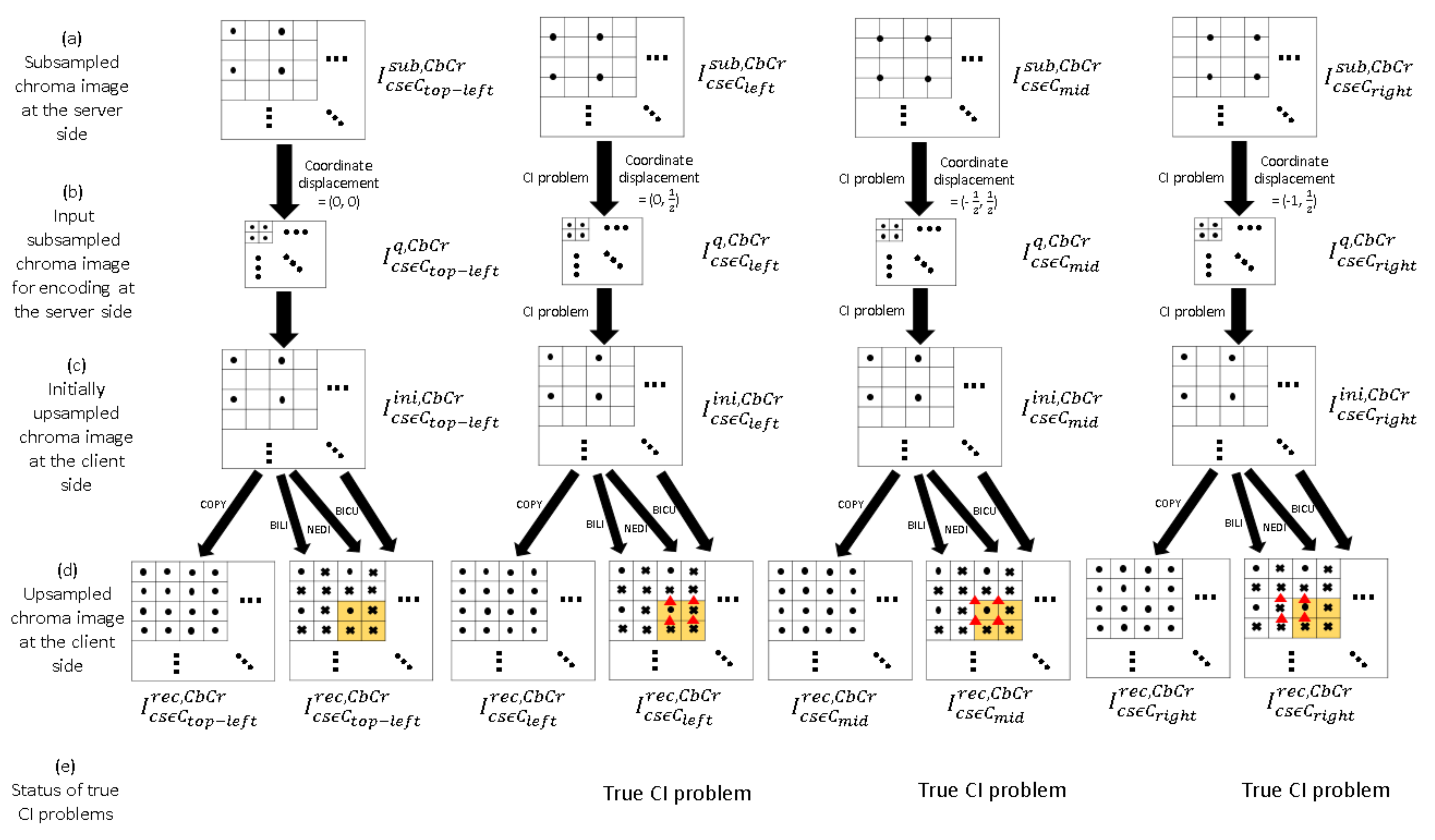}
\caption{The proposed flow diagram to analyze the CI problem for all combinations in {\bf CS}x{\bf CU}. (a) Subsampled chroma image. (b) Input subsampled chroma image for encoding. (c) Initially upsampled chroma image. (d) Upsampled chroma image. (e) Status of true CI problems.}
\label{pic:coordinate-inconsistency problem for four classes}
\end{figure*}

{\it 1) The proposed flow diagram to analyze the true CI problem and to calculate the coordinate displacement:} After performing one chroma subsampling $cs$ $\in$ {\bf CS} on the chroma image $I^{CbCr}$, it yields a subsampled chroma image $I^{sub,CbCr}_{cs}$, which can be expressed as $I^{sub,CbCr}_{cs \in C_{top-left}}$, $I^{sub,CbCr}_{cs \in C_{left}}$, $I^{sub,CbCr}_{cs \in C_{mid}}$,
or $I^{sub,CbCr}_{cs \in C_{right}}$. The four possible subsampled chroma images are depicted in Fig. \ref{pic:coordinate-inconsistency problem for four classes}(a).
For compression, as depicted in Fig. \ref{pic:coordinate-inconsistency problem for four classes}(b), each subsampled chroma image $I^{sub,CbCr}_{cs}$ is further rearranged to a quarter-sized subsampled chroma image $I^{q,CbCr}_{cs}$ which is stored in an array data structure under an integer coordinate system.

However, at the server side, for compression, moving the subsampled chroma pair of each 2$\times$2 chroma block from the subsampled chroma position ($\in$ {\bf SCP}) to the new position, namely (0, 1), often causes a coordinate displacement problem. Generally, as depicted by the four arrows between Fig. \ref{pic:coordinate-inconsistency problem for four classes}(a)
and Fig. \ref{pic:coordinate-inconsistency problem for four classes}(b), the corresponding four coordinate displacements equal (0, 0) (= (0, 1) - (0, 1)), (0, $\frac{1}{2}$)
(= (0, 1) - (0, $\frac{1}{2}$)), (-$\frac{1}{2}$, $\frac{1}{2}$) (= (0, 1) - ($\frac{1}{2}$, $\frac{1}{2}$)),
and (-1, $\frac{1}{2}$) (= (0, 1) - (1, $\frac{1}{2}$))
corresponding to $I^{q,CbCr}_{cs \in C_{top-left}}, I^{q,CbCr}_{cs \in C_{left}}, I^{q,CbCr}_{cs \in C_{mid}}, and I^{q,CbCr}_{cs \in C_{right}}$,
respectively. We conclude that for one chroma subsampling method $cs \in C_{left} \cup C_{mid} \cup C_{right}$, at the server side, preparing the quarter-sized subsampled chroma image $I^{q,CbCr}_{cs}$ for compression causes a CI problem due to its nonzero coordinate displacement (NCD) in the set {\bf NCD} = {(0, $\frac{1}{2}$), (-$\frac{1}{2}$, $\frac{1}{2}$), (-1, $\frac{1}{2}$)}.

After receiving the compressed quarter-sized subsampled chroma image $I^{q,CbCr}_{cs}$ by the decoder at the client side, each subsampled chroma pixel $I^{q,CbCr}_{cs}(i, j)$ is moved to $I^{rec,CbCr}_{cs}(2i, 2j)$, where $I^{rec,CbCr}_{cs}$ indicates the upsampled chroma image, for constructing the initially upsampled chroma image $I^{ini,CbCr}_{cs}$,
as depicted in Fig. \ref{pic:coordinate-inconsistency problem for four classes}(c).
Because from Fig. \ref{pic:coordinate-inconsistency problem for four classes}(a)
to Fig. \ref{pic:coordinate-inconsistency problem for four classes}(b), for $cs \in C_{left} \cup C_{mid} \cup C_{right}$, it causes a CI problem in the subsampled chroma image $I^{q,CbCr}_{cs}$,
the initially upsampled chroma image $I^{ini,CbCr}_{cs}$
in Fig. \ref{pic:coordinate-inconsistency problem for four classes}(c) thus inherits the CI
problem in Fig. \ref{pic:coordinate-inconsistency problem for four classes}(b) and the associated nonzero coordinate displacement in {\bf NCD}.

Further, all missing chroma pixels in $I^{ini,CbCr}_{cs}$ of Fig. \ref{pic:coordinate-inconsistency problem for four classes}(c) are reconstructed using the adopted chroma upsampling process $`cu'$ in {\bf CU}.
We first consider the chroma upsampling process `COPY'. After performing the COPY-based upsampling process on each initially subsampled chroma image in Fig. \ref{pic:coordinate-inconsistency problem for four classes}(c),
the reconstructed chroma pixels are denoted by black bullets of the upsampled chroma image in Fig. \ref{pic:coordinate-inconsistency problem for four classes}(d).
For each 2$\times$2 chroma block in $I^{ini,CbCr}_{cs}$, as depicted in Fig. \ref{pic:coordinate-inconsistency problem for four classes}(d), the three missing chroma pixels are reconstructed by copying the top-left subsampled chroma-pair of that block.
According to the analysis from Fig. \ref{pic:coordinate-inconsistency problem for four classes}(a) to Fig. \ref{pic:coordinate-inconsistency problem for four classes}(d),
we conclude that for any combination `$cs$-COPY', where `cs' is in {\bf CS}, no true CI problem occurs.

Next, we consider the chroma upsampling process `$cu$' $\in$ \{BILI, NEDI, BICU\}. After performing the upsampling process `$cu$' on each initially subsampled chroma image in Fig. \ref{pic:coordinate-inconsistency problem for four classes}(c),
the reconstructed chroma pixels are denoted by black cross-marked symbols of the reconstructed chroma image in Fig. \ref{pic:coordinate-inconsistency problem for four classes}(d).
In Fig. \ref{pic:coordinate-inconsistency problem for four classes}(d), each missing chroma pixel in $I^{ini,CbCr}_{cs}$ is reconstructed by the upsampling process `$cu$' referring to the neighboring chroma pairs of that missing chroma pixel. Consequently,
we conclude that for any combination in \{$C_{left}$, $C_{mid}$, $C_{right}$\}x\{BILI, NEDI, BICU\}, a true CI problem occurs eventually, as depicted in Fig. \ref{pic:coordinate-inconsistency problem for four classes}(e).
On the other hand, among the sixteen combinations in {\bf CS}x{\bf CU}, the true CI problems occur in only nine combinations in \{$C_{left}$, $C_{mid}$, $C_{right}$\}x\{BILI, NEDI, BICU\}.

From the above analysis of the true CI problems, we find that there are only three distinct coordinate displacements, namely (0, $\frac{1}{2}$), (-$\frac{1}{2}$, $\frac{1}{2}$), and (-1, $\frac{1}{2}$), corresponding to the three combinations in $C_{left}$x\{BILI, NEDI, BICU\}, the three combinations in $C_{mid}$x\{BILI, NEDI, BICU\},
and the three combinations in $C_{right}$x\{BILI, NEDI, BICU\}, respectively. Due to the nonzero coordinate displacement problem,
for each 2$\times$2 upsampled chroma block in Fig. \ref{pic:coordinate-inconsistency problem for four classes}(d), the four temporarily upsampled chroma pairs should be replaced by the four correct chroma pairs,
as depicted by the four red triangles in Fig. \ref{pic:coordinate-inconsistency problem for four classes}(d). The detailed re-interpolation based recovery strategy in our RCRI approach will be presented in Section IV.B.

{\it 2) Why the true CI problem degrades the quality of the reconstructed image:} Because for any combination, at the server side, we do not subsample the luma image $I^{Y}$ at all, the luma image has no CI problem from the chroma subsampling step to the chroma upsampling step at the client side. For any combination in \{$C_{left}$, $C_{mid}$, $C_{right}$\}x\{BILI, NEDI, BICU\}, we know a true CI problem occurs in the reconstructed chroma image $I^{rec,CbCr}$. Consequently, at the client side, the luma pixel $I^Y(i, j)$ and the upsampled chroma pixel $I^{rec,CbCr}(i, j)$ lead to a coordinate displacement problem each other. Finally, after converting the upsampled YCbCr image to the reconstructed RGB full-color image (or the reconstructed Bayer CFA image) by Eq. (\ref{eq:YCbCr2RGB}), it degrades the quality of the reconstructed image.


\section{The Proposed REDUCED CODEBOOK AND RE-INTERPOLATION (RCRI) APPROACH TO SOLVE THE UPI AND CI Problems}\label{sec:IIB}
To solve the UPI problem, in our RCRI (reduced codebook and re-interpolation) approach, we first build up a reduced codebook in which each codeword occupies four bits. For each codeword, the first two bits represent the chroma subsampling class instead of the chroma subsampling method used at the server side, and the last two bits represent the future chroma upsampling process preferred at the client side. This is why we call it the reduced codebook. Later, a (7, 4)-Hamming code is proposed to increase the robustness against the communication interference. To solve the true CI problem, in our RCRI approach, our coordinate displacement-based re-interpolation strategy will be presented in Subsection IV.B.

\subsection{The Reduced Codebook Design in Our RCRI Approach}\label{sec:IIIA}
From Fig. \ref{pic:position2}, we know there are four chroma subsampling classes, namely $C_{top-left}$, $C_{left}$, $C_{mid}$, and $C_{right}$, in {\bf CS} and there are four chroma upsampling processes, namely COPY, BILI, NEDI, and BICU, in {\bf CU}. To record the necessary information of each combination in {\bf CS}x{\bf CU}, a reduced 4-bit codebook is depicted in Table \ref{Tab:CSCxCU}.
In Table \ref{Tab:CSCxCU}, for each codeword, the first two bits are used to represent the chroma subsampling class, which corresponds to the chroma subsampling method used at the server side, and the last two bits are used to record the future chroma upsampling process preferred at the client side. For example,
the combination `JCDU-BICU' corresponding to `$C_{top-left}$-BICU' is expressed as the 4-bit codeword `0011'.

\begin{table}[htbp]
  \caption{THE REDUCED CODEBOOK USED FOR REPRESENTING THE CONSIDERED COMBINATIONS.}
  \label{Tab:CSCxCU}
  \centering
  \scalebox{0.9}{
  \begin{tabular}{|c|c|c|c|c|}
    \hline
    {} & COPY & BILI & NEDI & BICU \\
    \hline
    \multirow{2}{*}{$C_{top-left}$} & \multirow{2}{*}{0000} & \multirow{2}{*}{0001} & \multirow{2}{*}{0010} & \multirow{2}{*}{0011} \\
    & & & & \\
    \hline
    \multirow{2}{*}{$C_{left}$} & \multirow{2}{*}{0100} & \multirow{2}{*}{0101} & \multirow{2}{*}{0110} & \multirow{2}{*}{0111} \\
    & & & & \\
    \hline
    \multirow{2}{*}{$C_{mid}$} & \multirow{2}{*}{1000} & \multirow{2}{*}{1001} & \multirow{2}{*}{1010} & \multirow{2}{*}{1011} \\
    & & & & \\
    \hline
    \multirow{2}{*}{$C_{right}$} & \multirow{2}{*}{1100} & \multirow{2}{*}{1101} & \multirow{2}{*}{1110} & \multirow{2}{*}{1111} \\
    & & & & \\
    \hline
    \end{tabular}}
  \end{table}

To increase the robustness to tolerate one bit error against the communication interference, we deploy three redundant bits in the 4-bit codeword in Table \ref{Tab:CSCxCU} to form a (7, 4)-Hamming code \cite{Hamming}. Let each 4-bit codeword in Table \ref{Tab:CSCxCU} be denoted by `$m_1m_2m_3m_4$'.
Using the error correcting code technique, the corresponding (7, 4)-Hamming code is represented as a 7-bit codeword “$r_1r_2m_1r_3m_2m_3m_4$” in which the redundant 3-bit `$r_1r_2r_3$' is used for correcting the one bit error. For easy exposition,
let $p_1p_2p_3p_4p_5p_6p_7$ = $r_1r_2m_1r_3m_2m_3m_4$.
Based on the three equations with even parity:
$p_1 + p_2 + p_4 + p_5 = 0 \pmod{2}$,
$p_2 + p_3 + p_6 + p_7 = 0 \pmod{2}$,
and $p_1 + p_3 + p_5 + p_6 =0 \pmod{2}$, the corrupted one bit can be detected and corrected.

For example, suppose the decoder received the 7-bit error correction code, namely $p_1p_2p_3p_4p_5p_6p_7$ = 1100000. From the equality: $p_1p_2p_3p_4p_5p_6p_7$ = $r_1r_2m_1r_3m_2m_3m_4$, it yields `$m_1m_2m_3m_4$ = 1000'.
Based on three even parity equations, it yields
$p_1 + p_2 + p_4 + p_5 = 0 \pmod{2}$,
$p_2 + p_3 + p_6 + p_7 = 1 \pmod{2}$,
and $p_1 + p_3 + p_5 + p_6 = 1 \pmod{2}$. From the resultant 3-bit `011' calculated from the above three even parity equations, we know that the value of $p_3$ is corrupted. Therefore, the value of
$m_1$ is corrected from 0 to 1. Equivalently, the corrected 7-bit code should be 1110000, and the corrected 4-bit codeword equals 0001. Taking the first two bits of the corrected 4-bit codeword, namely 00, as a key to query Table \ref{Tab:CSCxCU} at the client side, the chroma subsampling class used at the server side is reported as $C_{top-left}$. Taking the last two bits of the corrected 4-bit codeword, namely 01, by Table \ref{Tab:CSCxCU}, the future chroma upsampling process preferred at the client side is reported as BILI. Table \ref{Tab:(7,4) Hamming code} illustrates the error correcting (7, 4)-Hamming codebook used in our RCRI approach. Consequently, for any combination in {\bf CS}x{\bf CU}, Table \ref{Tab:(7,4) Hamming code} can be used to solve the UPI problem in a robust way.
Table \ref{Tab:(7,4) Hamming code} is kept by the server side and the client side simultaneously.

\begin{table}[htbp]
  \caption{THE (7, 4)-HAMMING CODEBOOK USED IN OUR RCRI APPROACH.}
  \label{Tab:(7,4) Hamming code}
  \centering
  \scalebox{0.9}{
  \begin{tabular}{|c|c|c|c|c|}
    \hline
    {} & COPY & BILI & NEDI & BICU \\
    \hline
    \multirow{2}{*}{$C_{top-left}$} & \multirow{2}{*}{0000000} & \multirow{2}{*}{1110000} & \multirow{2}{*}{1001100} & \multirow{2}{*}{0111100} \\
    & & & & \\
    \hline
    \multirow{2}{*}{$C_{left}$} & \multirow{2}{*}{0101010} & \multirow{2}{*}{1011010} & \multirow{2}{*}{1100110} & \multirow{2}{*}{0010110} \\
    & & & & \\
    \hline
    \multirow{2}{*}{$C_{mid}$} & \multirow{2}{*}{1101001} & \multirow{2}{*}{0011001} & \multirow{2}{*}{0100101} & \multirow{2}{*}{1010101} \\
    & & & & \\
    \hline
    \multirow{2}{*}{$C_{right}$} & \multirow{2}{*}{1000011} & \multirow{2}{*}{0110011} & \multirow{2}{*}{0001111} & \multirow{2}{*}{1111111} \\
    & & & & \\
    \hline
    \end{tabular}}
  \end{table}

\subsection{The Re-Interpolation Strategy for Correcting the Deviationly Upsampled Chroma Image}\label{sec:IIIB}
We first consider the three combinations in $C_{left}$x\{BILI, NEDI, BICU\}. As depicted in Fig. \ref{pic:coordinate-inconsistency problem for four classes}, we know their common nonzero coordinate displacement vector is (0, $\frac{1}{2}$).
To solve the true CI problem which occurred in the three combinations, based on the coordinate displacement vector (0, $\frac{1}{2}$), for each 2$\times$2 upsampled chroma block in the temporarily reconstructed chroma image, the four temporarily upsampled chroma pairs, which are denoted by one black bullet and three cross-marked black bullets in Fig. \ref{pic:coordinate-inconsistency problem for four classes}(d),
are replaced by re-interpolating the four chroma pairs marked by the four red triangles in $I^{rec,CbCr}_{cs \in C_{left}}$ of Fig. \ref{pic:coordinate-inconsistency problem for four classes}(d).

Next, we consider the three combinations in $C_{mid}$x\{BILI, NEDI, BICU\}. As depicted in Fig. \ref{pic:coordinate-inconsistency problem for four classes}, we know their common nonzero coordinate displacement vector is (-$\frac{1}{2}$, $\frac{1}{2}$).
To solve the true CI problem which occurred in the three combinations, based on the coordinate displacement vector (-$\frac{1}{2}$, $\frac{1}{2}$), for each 2$\times$2 upsampled chroma block in the temporarily reconstructed
chroma image $I^{rec,CbCr}_{cs\in C_{mid}}$ of Fig. \ref{pic:coordinate-inconsistency problem for four classes}(d), the four temporarily upsampled chroma pairs are replaced by re-interpolating the four chroma pairs marked by the four red triangles in $I^{rec,CbCr}_{cs\in C_{mid}}$.

By the same argument, for each 2$\times$2 upsampled chroma block in $I^{rec,CbCr}_{cs\in C_{right}}$ of Fig. \ref{pic:coordinate-inconsistency problem for four classes}(d), the four temporarily upsampled chroma pairs can be recovered by re-interpolating the four chroma pairs marked by the four red triangles in $I^{rec,CbCr}_{cs\in C_{right}}$.
Consequently, the deviationly upsampled chroma image occurred in the nine combinations (= \{$C_{left}$, $C_{mid}$, $C_{right}$\}x\{BILI, NEDI, BICU\}) can be recovered using our coordinate displacement-based re-interpolation strategy.

\begin{table*}[htbp]
  \centering
  \caption{QUALITY ENHANCEMENT EFFECTS OF OUR RCRI APPROACH AGAINST THE UPI AND TRUE CI PROBLEMS FOR $I^{RGB}.$}
  \label{Tab:experimental result for CC}
\scalebox{0.55}{
\begin{tabular}{|c|c|c|c|c|c|c|c|c|c|c|c|c|c|c|c|c|c|c|c|c|}
  \hline
  \multirow{2}{*}{$I^{RGB}$} & \multicolumn{4}{c|}{IDID \cite{Y. Zhang}} & \multicolumn{4}{c|}{JCDU \cite{S. Wang}} & \multicolumn{4}{c|}{4:2:0(L)} & \multicolumn{4}{c|}{4:2:0(R)} & \multicolumn{4}{c|}{CSLM \cite{K. L. Chung-2020(rgb)}} \\ \cline{2-21}
  & COPY & BILI & NEDI \cite{Y. Zhang} & BICU & COPY & BILI & NEDI & BICU \cite{S. Wang} & COPY & BILI & NEDI & BICU & COPY & BILI & NEDI & BICU & COPY & BILI \cite{K. L. Chung-2020(rgb)} & NEDI & BICU \\ \hline
  \multirow{2}{*}{CPSNR} & 40.0832 & 45.2248 & 45.2343 & 44.5094 & 40.7675 & 45.3701 & 45.0005 & 45.6042 & 41.7039 & 43.6801 & 43.4977 & 43.6427 & 41.6744 & 40.6781 & 40.6171 & 40.5475 & 41.2692 & 42.2481 & 42.1925 & 41.3915 \\
                         & (45.2343) & (45.2343) & (45.2343) & (45.2343) & (45.6042) & (45.6042) & (45.6042) & (45.6042) & (45.1837) & (45.1837) & (45.1837) & (45.1837) & (44.4567) & (44.4567) & (44.4567) & (44.4567) & (46.2224) & (46.2224) & (46.2224) & (46.2224) \\
                         \hline
  Average                &
  \multicolumn{4}{c|}{\multirow{2}{*}{1.4715}} &
  \multicolumn{4}{c|}{\multirow{2}{*}{1.4186}} &
  \multicolumn{4}{c|}{\multirow{2}{*}{2.0526}} &                  \multicolumn{4}{c|}{\multirow{2}{*}{3.5774}} &
  \multicolumn{4}{c|}{\multirow{2}{*}{4.4471}} \\
  CPSNR gain             &
  \multicolumn{4}{c|}{}  &
  \multicolumn{4}{c|}{}  &
  \multicolumn{4}{c|}{}  &
  \multicolumn{4}{c|}{}  &
  \multicolumn{4}{c|}{}    \\ \hline
  \multirow{2}{*}{SSIMc} & 0.9710 & 0.9863 & 0.9865 & 0.9855 & 0.9742 & 0.9864 & 0.9858 & 0.9873 & 0.9779 & 0.9838 & 0.9834 & 0.9840 & 0.9778 & 0.9742 & 0.9741 & 0.9737 & 0.9750 & 0.9792 & 0.9792 & 0.9762 \\
                         & (0.9865) & (0.9865) & (0.9865) & (0.9865) & (0.9873) & (0.9873)& (0.9873) & (0.9873) & (0.9870) & (0.9870) & (0.9870) & (0.9870) & (0.9860) & (0.9860) & (0.9860) & (0.9860) & (0.9896) & (0.9896) & (0.9896) & (0.9896) \\
                         \hline
Average                &
\multicolumn{4}{c|}{\multirow{2}{*}{0.0042}} &
\multicolumn{4}{c|}{\multirow{2}{*}{0.0039}} &
\multicolumn{4}{c|}{\multirow{2}{*}{0.0047}} &                  \multicolumn{4}{c|}{\multirow{2}{*}{0.0111}} &
\multicolumn{4}{c|}{\multirow{2}{*}{0.0122}} \\
SSIMc gain             &
\multicolumn{4}{c|}{}  &
\multicolumn{4}{c|}{}  &
\multicolumn{4}{c|}{}  &
\multicolumn{4}{c|}{}  &
\multicolumn{4}{c|}{}    \\ \hline
  \multirow{2}{*}{FSIMc} & 0.9992 & 0.9996 & 0.9997 & 0.9996 & 0.9993 & 0.9997 & 0.9996 & 0.9997 & 0.9995 & 0.9996 & 0.9995 & 0.9996 & 0.9995 & 0.9992 & 0.9992 & 0.9992 & 0.9993 & 0.9995 & 0.9994 & 0.9993 \\
                         & (0.9997) & (0.9997) & (0.9997) & (0.9997) & (0.9997) & (0.9997) & (0.9997) & (0.9997) & (0.9997) & (0.9997) & (0.9997) & (0.9997) & (0.9997) & (0.9997) & (0.9997) & (0.9997) & (0.9997) & (0.9997) & (0.9997) & (0.9997) \\
                         \hline
Average                &
\multicolumn{4}{c|}{\multirow{2}{*}{0.0002}} &
\multicolumn{4}{c|}{\multirow{2}{*}{0.0001}} &
\multicolumn{4}{c|}{\multirow{2}{*}{0.0002}} &                  \multicolumn{4}{c|}{\multirow{2}{*}{0.0004}} &
\multicolumn{4}{c|}{\multirow{2}{*}{0.0003}} \\
FSIMc gain             &
\multicolumn{4}{c|}{}  &
\multicolumn{4}{c|}{}  &
\multicolumn{4}{c|}{}  &
\multicolumn{4}{c|}{}  &
\multicolumn{4}{c|}{}    \\ \hline
\end{tabular}}
\end{table*}

\section{Experimental Results}\label{sec:IV}
Based on the Kodak, IMAX, and Video datasets, and under the newly released versatile video coding (VVC) platform VTM-9.0 \cite{VVC-2019} for QP = 0, the thorough experimental results are illustrated to justify the quality enhancement effects of our RCRI approach against the CI and UPI problems for the traditional and state-of-the-art chroma subsampling methods.

All the concerned combinations are implemented on a computer with an Intel Core i7-7700 CPU 3.6 GHz and 24 GB RAM. The operating system is the Microsoft Windows 10 64-bit operating system. The program development environment is Visual C++ 2017.

We adopt the quality metrics, namely CPSNR (color peak signal-to-noise ratio), PSNR, SSIM (structure similarity index) \cite{Z. Wang-2004}, and FSIM (feature similarity index) \cite{L. Zhang}, to illustrate the quality enhancement effects of our RCRI approach against  the UPI and CI problems existing in the traditional and state-of-the-art combinations. The related quality metrics are defined below.

CPSNR is used to evaluate the average quality of the reconstructed RGB full-color images for one dataset with N images, and it is defined by

\begin{equation}
  \label{eq:CPSNR}
  \text{CPSNR}=\frac{1}{N}\sum_{n=1}^{N}10\log_{10}\frac{255^2}{CMSE}
\end{equation}
with $CMSE=\frac{1}{3WH}\sum_{p\in P}\sum_{c\in\{R,G,B\}}[I_{n,c}^{RGB}(p)-I_{n,c}^{RGB}(p)]^2$ in which P
$=\{(x,y) | 1 \leq x \leq H, 1 \leq y \leq W\}$ denotes the set of pixel coordinates in one $W \times H$ image. Here, N = 24, N = 18, and N = 200 for the Kodak, IMAX, and Video datasets, respectively. $I_{n,c}^{RGB}(p)$ and $I_{n,c}^{RGB}(p)$ denote the c-color value of the pixel at position p in the $n$th original RGB full-color image and the reconstructed one, respectively. In our experience, for fairness, each image in the Kodak dataset is downsampled to a quarter-sized one such that the average size of the downsampled images is close to the average size of the images in the IMAX dataset. The average CPSNR value equals the mean of the three CPSNR values for the three datasets. Similarly, the average PSNR value is used to evaluate the quality of the reconstructed Bayer CFA images for the three datasets.

For $I^{Bayer}$, SSIM \cite{Z. Wang-2004} is used to measure the joint preservation effects of luminance, contrast, and structure similarity between the original Bayer CFA image and the reconstructed one. For $I^{RGB}$, the SSIMc value is measured by the mean of the three SSIM values for the R, G, and B color planes.

For $I^{Bayer}$, FSIM \cite{L. Zhang} is an image quality metric with high consistency with the subjective evaluation. FSIM first utilizes the primary feature “phase congruency (PC)” and the minor feature “gradient magnitude” to obtain the local quality map, and then FSIM utilizes PC as a weighting function to obtain a quality score. For $I^{RGB}$, the FSIMc value is measured by the mean of the three FSIM values for the R, G, and B color planes.


\subsection{Quality Enhancement Merit of Our RCRI Approach for $I^{RGB}$
}\label{sec:IVA}
For $I^{RGB}$, this subsection presents the quality enhancement effects of our RCRI approach against the UPI and true CI problems for the traditional and state-of-the-art
combinations \cite{Y. Zhang}, \cite{S. Wang}, and \cite{K. L. Chung-2020(rgb)}. In the five traditional chroma subsampling methods, 4:2:0(L) and 4:2:0(R) are selected to balance our discussion of the experiments. In addition, the quality enhancement effect of our RCRI approach to the state-of-the-art combination \cite{Lin-2019} is also investigated.

Table \ref{Tab:experimental result for CC} illustrates the CPSNR gains of the reconstructed RGB full-color images using our RCRI approach against the UPI and true CI problems existing in the concerned combinations. For clarifying the quality enhancement effect of our RCRI approach, the CPSNR value of the reconstructed RGB full-color images using our RCRI approach for each combination is tabulated in the parenthesis `()'.

After deploying our RCRI approach in IDIDx{\bf CU}, except for IDID-NEDI \cite{Y. Zhang} without UPI and true CI problems, the CPSNR gains are 5.1511 (= 45.2343 - 40.0832 ) dB, 0.01 (= 45.2343 - 45.2248) dB, and 0.7249 dB w.r.t. IDID-COPY, IDID-BILI, and IDID-BICU, respectively. Suppose the probability of selecting each chroma upsampling process in {\bf CU} at the client side is the same and equals $\frac{1}{4}$.
For IDID-NEDI, the average CPSNR gain using our RCRI approach equals 1.4715 (= $\frac{1}{4}$(5.1511 + 0.01 + 0.7249)) dB, achieving a clear quality enhancement effect.

Similarly, after deploying our RCRI approach in JCDUx{\bf CU}, except for JCDU-BICU \cite{S. Wang} without UPI and true CI problems, the CPSNR gains using our RCRI approach are 4.8367 dB, 0.2341 dB, and 0.6037 dB w.r.t. JCDU-COPY, JCDU-BILI, and JCDU-NEDI, respectively; the average CPSNR gain equals 1.4186 (= $\frac{1}{4}$(4.8367 +  0.2341 + 0.6037)) dB, also achieving a clear quality enhancement effect.

After deploying our RCRI approach in 4:2:0(L)x{\bf CU}, 4:2:0(R)x{\bf CU}, and CSLMx{\bf CU}, the average CPSNR gains equal
2.0526 (= $\frac{1}{4}$(3.4798 + 1.5036 + 1.6860 + 1.5410)) dB, 3.5774 (= $\frac{1}{4}$(2.7823 + 3.7786 + 3.8396 + 3.9092)) dB, and 4.4471 (= $\frac{1}{4}$(4.9532 + 3.9743 + 4.0299 + 4.8309)) dB, respectively, also achieving significant quality enhancement effects.

In addition, our experimental results indicate that for $I^{RGB}$, after deploying our RCRI approach in `modified 4:2:0(A)'x({\bf CU} $\cup$ \{TN\}), except for ‘modified 4:2:0(A)’-TN \cite{Lin-2019}, the CPSNR gains for the five combinations are 2.1622 dB, 2.3352 dB, 2.5043 dB, and 2.2331 dB, respectively. Accordingly, the average CPSNR gain of our RCRI approach equals 1.84696 (= $\frac{1}{5}$(2.1622 + 2.3352 + 2.5043 + 2.2331)) dB, also achieving a clear quality enhancement effect.

Besides the CPSNR improvement, Table \ref{Tab:experimental result for CC} also demonstrates the $SSIM_c$ and $FSIM_c$ improvements of our RCRI approach for $I^{RGB}$. In fact, more chroma upsampling processes \cite{Zhang-2008}, \cite{J. Yang}, \cite{Z. Wang}, \cite{C. Dong}, \cite{T. Vermeir}, \cite{X. Wang} can be included in our study to justify the quality enhancement effect of the proposed RCRI approach.

\subsection{Quality Enhancement Merit of Our RCRI Approach for $I^{Bayer}$}\label{sec:IVA}

For $I^{Bayer}$, this subsection presents the quality enhancement effects of our RCRI approach against the UPI problems and true CI problems for 4:2:0(L), 4:2:0(R), and the state-of-the-art combinations \cite{Lin-2016}, \cite{K. L. Chung-2019}, \cite{K. L. Chung-2020(bayer)}, and \cite{Lin-2019} under the current coding environment.

From Table \ref{Tab:experimental result related to ABD}, we observe that after deploying our RCRI approach in
DMx{\bf CU}, GDx{\bf CU}, 4:2:0(L)x{\bf CU}, 4:2:0(R)x{\bf CU}, and $CSLM^{Bayer}$x{\bf CU}, the average PSNR gains equal
4.8134 (= $\frac{1}{4}$(5.2940 + 5.6031 + 8.3564)) dB, 5.7962 $\frac{1}{4}$(= (0.1047 + 7.5368 + 7.2515 + 8.2917)) dB, 2.4466 (= $\frac{1}{4}$(2.7829 + 2.3125 + 2.7198 + 1.9712)) dB, 4.1248 $\frac{1}{4}$(= (3.2360 + 4.3958 + 4.4658 + 4.4017)) dB, and
10.1278 (= $\frac{1}{4}$(9.8250 + 10.0740 + 9.7277 + 10.8843)) dB, respectively, achieving significant quality enhancement effects.

In addition, our experimental results indicate that for $I^{Bayer}$, after deploying our RCRI approach in `modified 4:2:0(A)'x({\bf CU} $\cup$ \{TN\}), except for `modified 4:2:0(A)'-TN \cite{Lin-2019}, the CPSNR gains are 2.0860 dB, 3.1487 dB, 3.5034 dB, and 2.8881 dB, respectively. Accordingly, the average PSNR gain of our RCRI approach
equals 2.3252 (= $\frac{1}{5}$(2.0860 + 3.1487 + 3.5034 + 2.8881)) dB, also achieving a clear quality enhancement effect.

Besides the PSNR improvement, Table \ref{Tab:experimental result related to ABD} also demonstrates the SSIM and FSIM improvements of our RCRI approach for $I^{Bayer}$.

\begin{table*}[htbp]
  \centering
  \caption{QUALITY ENHANCEMENT EFFECTS OF OUR RCRI APPROACH AGAINST THE UPI AND TRUE CI PROBLEMS FOR $I^{BAYER}$.}
  \label{Tab:experimental result related to ABD}
  \scalebox{0.55}{
  \begin{tabular}{|c|c|c|c|c|c|c|c|c|c|c|c|c|c|c|c|c|c|c|c|c|}
    \hline
    \multirow{2}{*}{$I^{Bayer}$} & \multicolumn{4}{c|}{DM \cite{Lin-2016}} &
    \multicolumn{4}{c|}{GD \cite{K. L. Chung-2019}} & \multicolumn{4}{c|}{4:2:0(L)} & \multicolumn{4}{c|}{4:2:0(R)} & \multicolumn{4}{c|}{$CSLM^{Bayer}$ \cite{K. L. Chung-2020(bayer)}} \\ \cline{2-21}
    & COPY \cite{Lin-2016} & BILI & NEDI & BICU & COPY & BILI \cite{K. L. Chung-2019} & NEDI & BICU & COPY & BILI & NEDI & BICU & COPY & BILI & NEDI & BICU & COPY & BILI \cite{K. L. Chung-2020(bayer)} & NEDI & BICU \\ \hline
    \multirow{2}{*}{PSNR} & 46.8980 & 41.6040 & 41.2949 & 41.5416 & 47.4129 & 40.9808 & 41.2661 & 40.2259 & 42.4150 & 42.8854 & 42.4781 & 43.2267 & 41.2777 & 40.1179 & 40.0479 & 40.1120 & 40.9136 & 40.6646 & 41.0109 & 39.8543 \\
                          & [46.8980] & [46.8980] & [46.8980] & [46.8980] & [48.5176] & [48.5176] & [48.5176] & [48.5176] & [45.1979] & [45.1979] & [45.1979] & [45.1979] & [44.5137] & [44.5137] & [44.5137] & [44.5137] & [50.7386] & [50.7386] & [50.7386] & [50.7386] \\
                          \hline
   Average                                 &
   \multicolumn{4}{c|}{\multirow{2}{*}{4.8134}} &
   \multicolumn{4}{c|}{\multirow{2}{*}{5.7962}} &
   \multicolumn{4}{c|}{\multirow{2}{*}{2.4466}} &                  \multicolumn{4}{c|}{\multirow{2}{*}{4.1248}} &
   \multicolumn{4}{c|}{\multirow{2}{*}{10.1278}} \\
   PSNR gain              &
   \multicolumn{4}{c|}{}  &
   \multicolumn{4}{c|}{}  &
   \multicolumn{4}{c|}{}  &
   \multicolumn{4}{c|}{}  &
   \multicolumn{4}{c|}{}    \\ \hline
    \multirow{2}{*}{SSIM} & 0.9981 & 0.9946 & 0.9943 & 0.9945 & 0.9983 & 0.9936 & 0.9942 & 0.9924 & 0.9955 & 0.9959 & 0.9955 & 0.9961 & 0.9943 & 0.9925 & 0.9924 & 0.9924 & 0.9937 & 0.9931 & 0.9938 & 0.9917 \\
                          & [0.9981] & [0.9981] & [0.9981] & [0.9981] & [0.9987] & [0.9987] & [0.9987] & [0.9987] & [0.9973] & [0.9973] & [0.9973] & [0.9973] & [0.9969] & [0.9969] & [0.9969] & [0.9969] & [0.9991] & [0.9991] & [0.9991] & [0.9991] \\
                          \hline
Average                                 &
\multicolumn{4}{c|}{\multirow{2}{*}{0.0027}} &
\multicolumn{4}{c|}{\multirow{2}{*}{0.0041}} &
\multicolumn{4}{c|}{\multirow{2}{*}{0.0016}} &                  \multicolumn{4}{c|}{\multirow{2}{*}{0.0040}} &
\multicolumn{4}{c|}{\multirow{2}{*}{0.0059}} \\
SSIM gain              &
\multicolumn{4}{c|}{}  &
\multicolumn{4}{c|}{}  &
\multicolumn{4}{c|}{}  &
\multicolumn{4}{c|}{}  &
\multicolumn{4}{c|}{}    \\ \hline
    \multirow{2}{*}{FSIM} & 0.9984 & 0.9975 & 0.9973 & 0.9975 & 0.9986 & 0.9969 & 0.9972 & 0.9965 & 0.9977 & 0.9981 & 0.9979 & 0.9982 & 0.9964 & 0.9967 & 0.9968 & 0.9968 & 0.9971 & 0.9968 & 0.9972 & 0.9964 \\
                          & [0.9984] & [0.9984] & [0.9984] & [0.9984] & [0.9992] & [0.9992] & [0.9992] & [0.9992] & [0.9987] & [0.9987] & [0.9987] & [0.9987] & [0.9983] & [0.9983] & [0.9983] & [0.9983] & [0.9996] & [0.9996] & [0.9996] & [0.9996] \\
                          \hline
Average                                 &
\multicolumn{4}{c|}{\multirow{2}{*}{0.0007}} &
\multicolumn{4}{c|}{\multirow{2}{*}{0.0019}} &
\multicolumn{4}{c|}{\multirow{2}{*}{0.0007}} &                  \multicolumn{4}{c|}{\multirow{2}{*}{0.0016}} &
\multicolumn{4}{c|}{\multirow{2}{*}{0.0027}} \\
FSIM gain              &
\multicolumn{4}{c|}{}  &
\multicolumn{4}{c|}{}  &
\multicolumn{4}{c|}{}  &
\multicolumn{4}{c|}{}  &
\multicolumn{4}{c|}{}    \\ \hline
  \end{tabular}}
  \end{table*}

\section{Conclusion}\label{sec:V}
For $I^{RGB}$ and $I^{Bayer}$, we have presented the proposed RCRI (reduced codebook and re-interpolation) approach to solve the UPI and true CI problems existing in the traditional and state-of-the-art combinations under the current coding environment. Based on the Kodak, IMAX, and Video datasets, the comprehensive experimental results have justified the clear quality enhancement effects after deploying our RCRI approach in the traditional and state-of-the-art combinations \cite{Y. Zhang}, \cite{S. Wang}, \cite{K. L. Chung-2020(rgb)},
\cite{Lin-2016}, \cite{K. L. Chung-2019}, \cite{Lin-2019}, \cite{K. L. Chung-2020(bayer)}.

Our future work is to integrate our RCRI approach, the discrete cosine transform (DCT) based subsampling method \cite{S. Y. Zhu}, and the DCT based quantization error-minimization method \cite{S. Zhu} to achieve better quality of the reconstructed images in JPEG \cite{JPEG}. Moreover, our additional future work is to apply our RCRI approach to enhance the accuracy of the cross-component linear model for
chroma component coding \cite{J. Li}, \cite{K. Zhang} and to solve the accuracy degradation problem existing in the  luma-guided winner-first voting strategy \cite{Chung-2017-2} at the client side for identifying the chroma subsampling method used at the server side for high QP values.

\section*{Acknowledgment}
The authors appreciate  the proofreading help of Ms. C. Harrington  to improve the manuscript.
\ifCLASSOPTIONcaptionsoff
  \newpage
\fi



%

%

\begin{IEEEbiography}[{\includegraphics[width=1in,height=1.25in,clip,keepaspectratio]{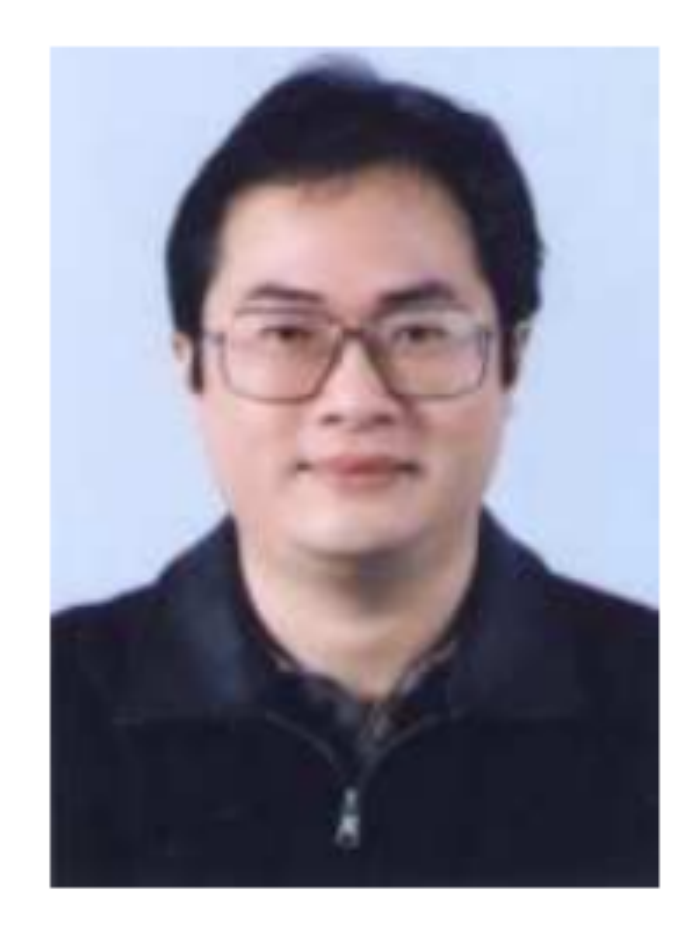}}]
    {Kuo-Liang Chung}
    (SM01)received his B.S., M.S., and Ph.D. degrees from National Taiwan University, Taipei, Taiwan in 1982, 1984, and 1990, respectively. He has been one Chair Professor of the Department of Computer Science and Information Engineering at National Taiwan University
  of Science and Technology, Taipei, Taiwan since 2009. He was the recipient of the Distinguished Research Award (2004-2007; 2019-2022) and Distinguished Research Project Award (2009-2012) from the Ministry of Science and Technology of Taiwan. In 2020, he received the K. T. Li Fellow Award from the Institute of Information Computing Machinery, Taiwan. He has been an Editor and Associate Editor of Signals and the Journal of Visual Communication and Image Representation since 2020 and 2011, respectively. His research interests include machine learning, image processing, and video compression.
  \end{IEEEbiography}

  \begin{IEEEbiography}[{\includegraphics[width=1in,height=1.25in,clip,keepaspectratio]{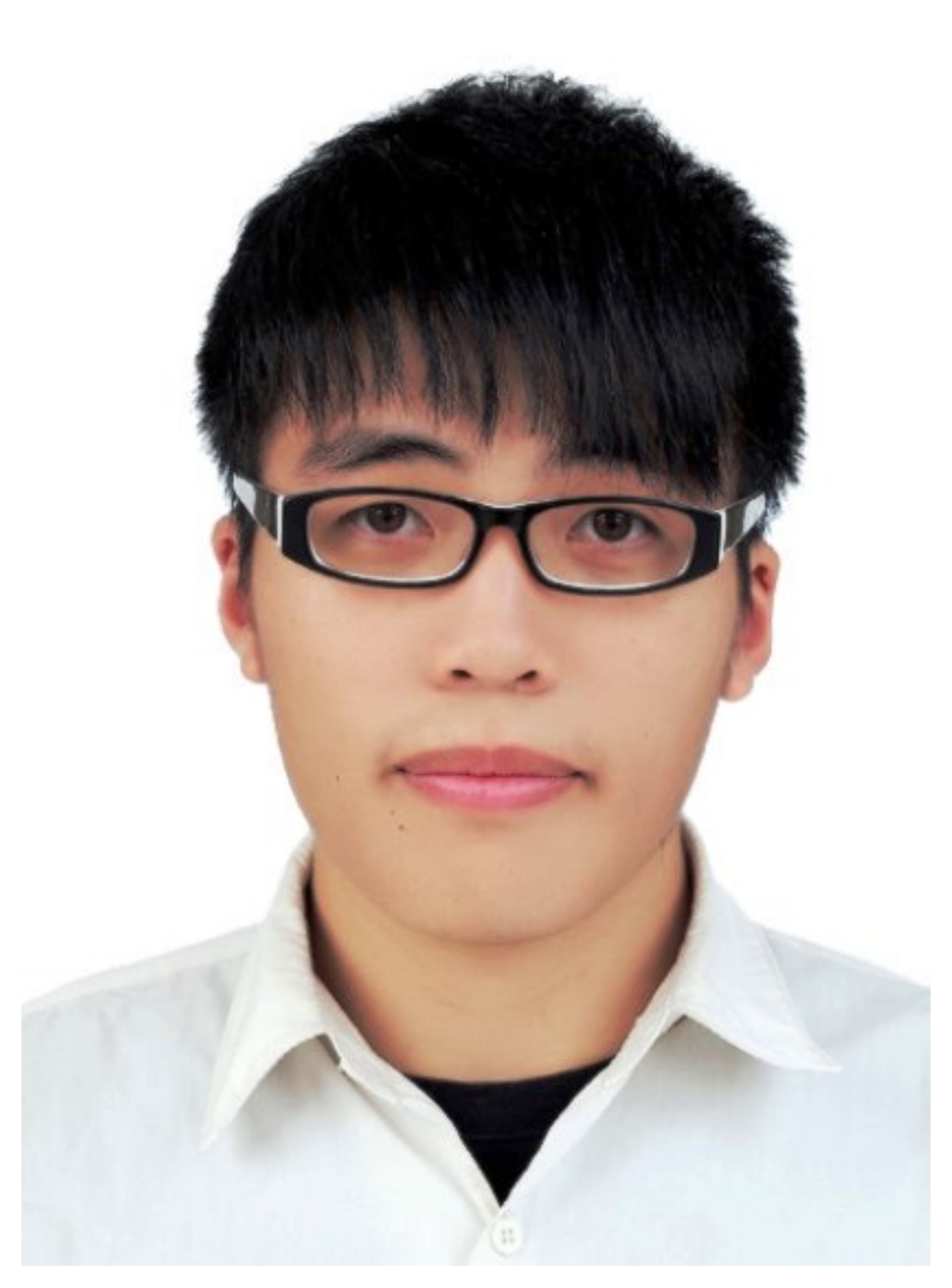}}]
    {Chen-Wei Kao}
    received his B.S. degree in Computer Science and Engineering from the National Taiwan Ocean University, Keelung, Taiwan, in 2019. He is currently working towards his M.S. degree in Computer Science and Information Engineering at the National Taiwan University of Science and Technology, Taipei, Taiwan. His research interests include image processing and video compression.
  \end{IEEEbiography}



\end{document}